\documentstyle[preprint,aps,floats]{revtex}
\input epsf.tex
\def\DESepsf(#1 width #2){\epsfxsize=#2 \epsfbox{#1}}

\def\etal{ {\em et al.}}

\def\be {\begin{equation}}
\def\ee {\end{equation}}
\def\barr{\begin{array}}
\def\earr{\end{array}}
\def\dis{\displaystyle}
\def\ra{\rightarrow}

\def\bra {\langle}
\def\ket{\rangle}

\def\l {\lambda}
\def\rp {R_p}
\def\rpv {R_p\!\!\!\!\!\!/~~}
\def\hlp {H_{eff}^{\lambda'}}

\def\betak {B\rightarrow \eta K}
\def\betak0 {B^{0}\rightarrow \eta' K^{0}}

\def\betapkstr0 {B^{0}\rightarrow \eta' K^{*0}}
\def\bpetak0 {B^{0}\rightarrow \eta' K^{0}}

\def\msnus {m_{\tilde\nu_{iL}}^2}
\def\msells {m_{\tilde e_{iL}}^2}

\def\lappeq{\mathrel{\rlap{\raise.5ex\hbox{$<$}}
                    {\lower.5ex\hbox{$\sim$}}}}

\begin{document}
\preprint{\vbox{\hbox{}\hbox{}\hbox{}}}
\draft
\title{A Consistent Resolution of Possible Anomalies \\
in $B^0 \to \phi K_S$ and $B^+ \to \eta' K^+$ Decays}

\author{ B. Dutta$^{1}$\footnote{duttabh@yogi.phys.uregina.ca}, ~~
C.~S. Kim$^2$\footnote{cskim@mail.yonsei.ac.kr,~~~
http://phya.yonsei.ac.kr/\~{}cskim/}~~ and~~ Sechul
Oh$^{2,3}$\footnote{scoh@post.kek.jp}  }

\address{
$^1$Department of Physics, University of Regina, SK, S4S 0A2, Canada\\
$^2$Department of Physics and IPAP, Yonsei University, Seoul
120-479, Korea\\
$^3$Theory Group, KEK, Tsukuba, Ibaraki 305-0801, Japan}
\maketitle

\begin{abstract}
In the framework of $R$-parity violating ($\rpv$) supersymmetry, we try to
find a \emph{consistent} explanation for \emph{both} recently measured CP
asymmetry in $B^0 \to \phi K_S$ decay \emph{and} the large branching ratio
of $B^{\pm}\rightarrow \eta' K^{\pm}$ decay, which are inconsistent with
the Standard Model (SM) prediction.
We also investigate other charmless hadronic $B \to PP$ and $B \to VP$ decay
modes whose experimental data favor the SM: for instance,
recently measured CP asymmetries in $B^0 \to \eta^{\prime} K_S$ and
$B^0 \to J /\Psi K_S$.
We find that all the observed data can be accommodated for certain values
of $\rpv$ couplings.
\end{abstract}

\newpage
The main mission of $B$ factories, such as KEK$-$B and SLAC$-$B,
is to test the Standard Model (SM) and further to find possible
new physics effects in $B$ meson systems. For this goal, a variety
of useful observables can be measured to be compared with
theoretical expectations.  One of the important observables is CP
asymmetries in various $B$ meson decays, and another is branching
ratios (BRs) for rare $B$ decay processes.

CP violation in $B$ system has been confirmed in measurements of
time-dependent CP asymmetries in $B \to J/ \Psi K_S$ decay. The
world average of the asymmetry in $B \to J/ \Psi K_S$ \cite{nir}
is given by
\begin{equation}
\sin(2\beta)_{J/ \Psi K_S} = 0.734 \pm 0.054~, \label{sin2bpsik}
\end{equation}
which is consistent with the SM expectation.
However, recent measurements of $\sin(2\beta)$ in $B^0 \to \phi K_S$ decay disagree
with the above value.
The measurements of time-dependent CP asymmetries in $B^0 \to \phi K_S$ have been
recently reported by BaBar \cite{babar} and Belle \cite{belle}, respectively:
\begin{eqnarray}
\sin(2\beta)_{\phi K_S}^{BaBar} &=& -0.19^{+0.52}_{-0.50} \pm 0.09 ~, \\
\sin(2\beta)_{\phi K_S}^{Belle} &=& -0.73 \pm 0.64 \pm 0.18~.
\end{eqnarray}
Both experimental results show (a tendency of) negative values of $\sin(2\beta)$.

Because within the SM the difference between the asymmetries in $B
\to J/ \Psi K_S$ and $B^0 \to \phi K_S$ is expected to be
\cite{diff}
\begin{equation}
| \sin(2\beta)_{J/ \Psi K_S} - \sin(2\beta)_{\phi K_S} | \lesssim
O(\lambda^2)
\label{diffsin2b}
\end{equation}
with $\lambda \approx 0.2$, these results indicate 2.7 $\sigma$
deviation from the SM prediction and may reveal new physics
effects.

The decay process $B^0 \to \phi K_S$ is a pure penguin process $b
\to s \bar s s$, and in the SM the time-dependent CP asymmetry of
this mode is expected to measure the same $\sin(2\beta)$ as in $B
\to J/ \Psi K_S$, as shown in Eq. (\ref{diffsin2b}). If there are
any new physics effects in the measurements of time-dependent CP
asymmetries, the effects can arise from new contributions to the
$B^0 - \bar B^0$ mixing amplitude and/or the decay amplitude of
each mode. Because the new physics effects to the $B^0 - \bar B^0$
mixing would be universal to all the neutral $B^0_d$ meson decays,
the discrepancy between the measured values of $\sin(2\beta)_{J/
\Psi K_S}$ and $\sin(2\beta)_{\phi K_S}$ may indicate the
possibility of new physics effects in the decay amplitude of $B^0
\to \phi K_S$. Supposing that the mode $B^0 \to \phi K_S$ reveals
new physics effects \cite{diff,ddo1}, one might expect that other
modes having the same internal quark level process $b \to s \bar s
s$ (e.g., $B^0 \to \eta' K_S$) would reveal the similar effects.
Interestingly, the recent measurements of CP asymmetries in $B^0
\to \eta' K_S$ by Belle \cite{belle} agree well with the results
of $\sin(2\beta)_{J/ \Psi K_S}$ given in Eq. (\ref{sin2bpsik})
(see Table I). Therefore, any successful explanation invoking new
physics should accommodate the data disfavoring the SM, such as
$\sin(2\beta)_{\phi K_S}$, and simultaneously all the data
favoring the SM, such as $\sin(2\beta)_{\eta' K_S}$ and
$\sin(2\beta)_{J/ \Psi K_S}$. Motivated by the recent measurements
of $\sin(2\beta)_{\phi K_S}$, a few first analyses
\cite{nir,hiller,datta,ciuchini,raidal} have been done. But, all
of them explain only the deviation of $\sin(2\beta)_{\phi K_S}$
from the SM prediction, or in addition, make (at most) qualitative
statements on $\sin(2\beta)_{\eta' K_S}$
\cite{nir,hiller,datta,ciuchini,raidal}.

There also exist plenty of experimental data observed for
charmless hadronic $B \to PP$ ($P$ denotes a pseudoscalar meson)
and $B \to VP$ ($V$ denotes a vector meson) decay modes, which are
well understood within the SM. However, among the $B\ra PP$ decay
modes, the BR of the decay mode $B^{\pm} \to \eta^{\prime}
K^{\pm}$ is found to be still larger than that expected  within
the SM\footnote{Attempts to explain the large ${\cal
B}(B^{\pm}\ra\eta'K^{\pm})$ within the SM have been made: e.g., by
using the anomalous $g-g-\eta'$ coupling \cite{as}.}
\cite{cleo,belle2,babar2,dko}: the experimental world average is
${\cal B}(B^{\pm}\ra\eta'K^{\pm})= (75 \pm 7) \times 10^{-6}$. The
SM contribution is about $3\sigma$ smaller than the experimental
world average \cite{dko}. Among the $B \to VP$ decay modes, the
experimentally observed BR for the decay $B^{0} \to \eta K^{*0}$
has been also found to be $2\sigma$ larger than the SM expectation
\cite{dko}.

In this letter we try to find a \emph{consistent} explanation for
\emph{all the observed data} in charmless hadronic $B \to PP$ and $B \to VP$ decays
in the framework of $R$-parity violating ($\rpv$) supersymmetry (SUSY).
In $\rpv$ SUSY, the effects of $\rpv$ couplings on $B$ decays can appear at tree level
and can be in some cases comparable to the SM contribution.
Our main focus is on
the recent measurement of CP asymmetry in $B^0 \to \phi K_S$ and the large BR for
$B^{\pm} \to \eta^{\prime} K^{\pm}$: both data appear to be inconsistent with
the SM prediction.
In order to achieve this goal, we investigate all the observed $B \to PP$ and
$B \to VP$ decay modes and \emph{explicitly} show that indeed all the observed data
can be accommodated in a consistent way for certain values of $\rpv$ couplings.

The $\rpv$ part of the superpotential of the minimal supersymmetric standard
model  (MSSM) can contain terms of the form
\be
 {\cal W}_{\rpv}=\kappa_iL_iH_2 + \l_{ijk}L_iL_jE_k^c + \l'_{ijk}L_iQ_jD_k^c
          + \l''_{ijk}U_i^cD_j^cD_k^c \,
      \label{superpot}
\ee
where $E_i$, $U_i$ and $D_i$ are respectively the $i$-th type of  lepton,
up-quark and down-quark singlet superfields, $L_i$ and
$Q_i$ are the SU$(2)_L$ doublet lepton and quark superfields, and
$H_2$ is the Higgs doublet with the appropriate hypercharge.   From the symmetry
reason, we need $\l_{ijk}=-\l_{jik}$ and
$\l''_{ijk}=-\l''_{ikj}$. The bilinear terms can be rotated away with
redefinition of lepton and  Higgs superfields, but the effect reappears as
$\l$s, $\l'$s and  lepton-number violating soft terms \cite{roy}.  The first
three terms of Eq. (\ref{superpot}) violate the lepton number, whereas the fourth
term violates the baryon number.  We do not want all these terms to be  present
simultaneously due to catastrophic rates for proton decay.   In order to prevent
proton decay, one set needs to be forbidden.

For our purpose, we will assume only  $\l'-$type  couplings to be present.
Then, the effective Hamiltonian for charmless hadronic $B$ decay can be
written  as\cite{prev} \be \barr{rcl} \dis \hlp (b\ra \bar d_j d_k d_n)
   & = & \dis d^R_{jkn} [ \bar d_{n\alpha} \gamma^\mu_L d_{j\beta}
                           \; \bar d_{k\beta} \gamma_{\mu R} b_{\alpha}]
            + d^L_{jkn} [ \bar d_{n\alpha} \gamma^\mu_L b_{\beta}
                           \; \bar d_{k\beta} \gamma_{\mu R} d_{j\alpha}]
              \ ,
             \\[1.5ex]
\dis \hlp (b\ra \bar u_j u_k d_n)
   & = & \dis u^R_{jkn} [\bar u_{k\alpha} \gamma^\mu_L u_{j\beta}
                           \; \bar d_{n\beta} \gamma_{\mu R} b_{\alpha}]
              \ ,
              \\[1,5ex]

\earr
    \label{rp_hamilt}
\ee
with
\be
 \barr{rclcrclcl} d^R_{jkn} &=& \dis
      \sum_{i=1}^3 {\l'_{ijk}\l'^{\ast}_{in3}\over 8\msnus},
      &  &  d^L_{jkn} &=& \dis \sum_{i=1}^3 {\l'_{i3k}\l'^{\ast}_{inj}\over
8\msnus},
      &  & (j,k,n=1,2)
           \\[1.5ex]  u^R_{jkn} &=& \dis \sum_{i=1}^3
{\l'_{ijn}\l'^{\ast}_{ik3}\over 8\msells},
      &  &
      &  &
      &  & (j,k=1, \ n=2),
\earr
\ee
where $\alpha$ and $\beta$ are color indices and
$\gamma^\mu_{R, L} \equiv \gamma^\mu (1 \pm \gamma_5)$. The leading order QCD
correction to this operator is given by a scaling factor $f\simeq 2$ for
$m_{\tilde\nu}=200$ GeV.

The available data on low energy processes can be used to impose rather
strict constraints on many of these  couplings
\cite{constraints,products,herbi}. Most such bounds have been  calculated
under the assumption of there being  only one non-zero $\rpv$ coupling.
There is no strong argument to have only one $\rpv$  coupling being
nonzero. In fact, a hierarchy of couplings may be  naturally obtained
\cite{products} on account of the mixings in either of the quark and squark
sectors. In this work, we try to find out the values of  such $\rpv$
couplings for which  all available data are simultaneously satisfied. An
important role will be played by the $\l'_{32i}$ -type couplings, the
constraints on which are relatively weak. The BR of $B \to X_s \nu\nu$ can
put bound on $\l'_{322}\l'^{\ast}_{323}$ in certain limits. Using Ref.
\cite{grossman} and the  experimental limit (BR$<6.4\times 10^{-4}$) on
the BR of $B \to X_s \nu\nu$ \cite{ALEPH}, we find that $\l'\leq 0.07 $
(for $m_{\tilde q}=200$ GeV).
However, if we go to any realistic scenario, for example grand
unified models (with $R$-parity violation), we find a natural hierarchy
among the sneutrino and squark masses. The squark masses are much heavier
than the sneutrino masses and the bound does not apply any more for
$m_{\tilde \nu}=200$ GeV.

In our calculation, we use the effective Hamiltonian and the effective Wilson
coefficients (WCs) for the processes $b\ra s\bar qq'$ and $b\ra d\bar qq'$ given
in Ref. \cite{ddo}.
The decay amplitude of $B^- \ra \phi K^-$ decay mode is given by
\be
\bar {\cal A}_{\phi K} = \bar {\cal A}^{SM}_{\phi K} + \bar {\cal A}^{\rpv}_{\phi K}~,
\ee
where the SM part of the amplitude is
\be
\bar {\cal A}^{SM}_{\phi K} = -{G_F \over \sqrt{2}} V_{tb} V^*_{ts} (a_3 +a_4 +a_5
-{1 \over 2} a_7 -{1 \over 2} a_9 -{1 \over 2} a_{10}) A_{\phi}~,
\ee
and the $\rpv$ part of the amplitude involves only
$d^L_{222}$ and  $d^R_{222}$:
\be
\barr{rcl}
\bar {\cal A}^{\rpv}_{\phi K} &=& \dis

          \left( d^L_{222} + d^R_{222} \right) \:
                 \left[  \xi A_{\phi}\right]~,
\label{rpvphik} \earr \ee with $A_{\phi}=\bra K|\bar s
\gamma^{\mu} (1-\gamma_5) b|B\ket \; \bra \phi|\bar s \gamma_{\mu}
s|0\ket$. This particular structure of $A_{\phi}$ is obtained from
the operators $\bar d_{n\alpha} \gamma^\mu_{L(R)} d_{j\beta} ~
\bar d_{k\beta} \gamma_{\mu R(L)} b_{\alpha}$ given in Eq.
(\ref{rp_hamilt}), which are derived from the operators
$\bar{s}_{L(R)} s_{R(L)} \bar{s}_{R(L)} b_{L(R)}$ by Fierz
transformation \cite{guetta}.  Here the effective coefficients
$a_i$ are defined as $a_i = c^{eff}_i + \xi c^{eff}_{i+1}$ ($i =$
odd) and $a_i = c^{eff}_i + \xi c^{eff}_{i-1}$ ($i =$ even) with
the effective WCs $c^{eff}_i$ at the scale $m_b$ \cite{ddo}, and
by treating $\xi\equiv 1/N_c$ ($N_c$ denotes the effective number
of color) as an adjustable parameter.

The CP asymmetry parameter $\sin(2 \tilde \beta)_{XY}$ for $B \to XY$
can be written as
\be
\sin(2 \tilde \beta)_{XY} = - {2~ {\rm Im} \lambda_{XY} \over
(1+ |\lambda_{XY}|^2)}~,
\ee
where $\lambda_{XY} = e^{- 2i\beta} (\bar {\cal A}_{XY}/ {\cal A}_{XY})
= e^{- i (2\beta + \theta)} |\bar {\cal A}_{XY}/ {\cal A}_{XY}|$
with $\bar {\cal A}_{XY}/ {\cal A}_{XY}
= e^{- i \theta} |\bar {\cal A}_{XY}/ {\cal A}_{XY}|$.
Here $\beta$ denotes the CP angle in the SM and $\tilde \beta$ denotes
the effective CP angle given by $2 \tilde \beta = 2 \beta + \theta$ with
a new weak phase $\theta$ arising from new physics.
${\cal A}_{XY} ={\cal A}^{SM}_{XY} +{\cal A}^{\rpv}_{XY}$ and $\bar {\cal A}_{XY}$
is its CP-conjugate amplitude.
For $B^0 \to \phi K_S$, in the SM, $\bar {\cal A}_{\phi K_S}/ {\cal A}_{\phi K_S} =1$,
so $\tilde \beta = \beta$.
But, in $\rpv$ SUSY, generally $|\bar {\cal A}_{\phi K_S}/ {\cal A}_{\phi K_S}| \neq 1$
and $\theta \neq 0$.

The $\rpv$ part of the amplitude of  $B^- \ra \eta' K^-$ decay is
\be
\barr{rcl}
\bar {\cal A}^{\rpv}_{\eta' K} &=& \dis
       \left( d^R_{121} - d^L_{112} \right) \xi A_{\eta'}^u
        +
          \left( d^L_{222} - d^R_{222} \right) \:
                 \left[  \frac{\bar m}{m_s}
                   \left(A_{\eta'}^s -A_{\eta'}^u\right)- \xi  A_{\eta'}^s\right]
        \\[1.5ex]
       & + &  \dis
        \left(d^L_{121} - d^R_{112} \right) \frac{\bar m}{m_d} A_{\eta'}^u
       +
       u^R_{112} \left[\xi A_{\eta'}^u - {2 m_K^2 A_K \over (m_s+m_u) (m_b-m_u)}
                 \right],
\label{rpvetak}
\earr \ee
where $\bar m \equiv m^2_{\eta'} /
(m_b-m_s)$ and
\[
\barr{rcl} A^{u(s)}_{\eta'} & = & f^{u(s)}_{\eta'} F^{B \to K} (m_B^2 -m_K^2)~,\\
A_{K} & = & f_K F^{B \to \eta'} (m_B^2 -m_{\eta'}^2) .\earr
\]
$F^{B \to K(\eta')}$ denotes the hadronic form factor for $B \to K
(\eta')$ and $f_{K(\eta')}$ is the decay constant of $K(\eta')$
meson\footnote{The definition of $f^{u(s)}_{\eta'}$ is given in
Ref. \cite{ddo}, considering the $\eta - \eta'$ mixing.  The
mixing angle $\theta = -22^0$ is used. }. Analogous expressions
hold for $B^{\pm}\ra\eta K^{\pm}$ where we have to replace
$A^u_{\eta^{\prime}}$ by $A^u_{\eta}$,
 $A^s_{\eta^{\prime}}$ by $A^s_{\eta}$ and $m_{\eta^{\prime}}$ by $m_{\eta}$.
Replacing a pseudoscalar meson by a vector meson, we
 also get similar expressions for the amplitudes of $B^{\pm(0)}\ra\eta'
K^{*\pm(0)}$ modes.

In Eqs. (\ref{rpvphik}) and (\ref{rpvetak}), we note that $\bar
{\cal A}^{\rpv}_{\phi K}$ includes the $(d^L_{222} +d^R_{222})$
term only, while $\bar {\cal A}^{\rpv}_{\eta' K}$ includes the
$(d^L_{222} -d^R_{222})$ term. Furthermore, in the $\rpv$ SUSY
framework, we find that the positive values of $d^R_{222}$ and
negative values of $d^L_{222}$ can increase the BR for the process
$B^{\pm}\ra\eta'K^{\pm}$, keeping most of the other $B\ra PP$ and
$B\ra VP$ modes unaffected. The other $\rpv$ combinations are
either not enough to increase in the BR for
$B^{\pm}\ra\eta'K^{\pm}$ or affect too many other modes. From now
on, we will concentrate on contributions of $d^L_{222}$ and
$d^R_{222}$ which are relevant to the process $b \to s \bar s s$.
We divide our results into the following two cases.

{\bf Case 1:} We use the following input parameters: the CP angle
$\gamma=110^0$, the $s$ quark mass $m_s~ ({\rm at}~ m_b ~{\rm
scale}) =85$ MeV, and the decay constants and the hadronic form
factors used in Ref. \cite{dko}. We take $\beta = 26^0$ as our
input, which corresponds to $\sin(2\beta)_{J/ \Psi K} =0.78$ given
in Eq. (\ref{sin2bpsik}). We set $d^L_{222} = k e^{-i \theta'}$
and $d^R_{222} = -k e^{i \theta'}$, where $k = |d^L_{222}| =
|d^R_{222}|$ and $\theta'$ is a new weak phase arising from
$d^L_{222} \propto \lambda^{\prime}_{i32} \lambda^{\prime
*}_{i22}$, or $(d^R_{222})^* \propto \lambda^{\prime *}_{i22}
\lambda^{\prime}_{i23}$. In this case, $\bar {\cal A}^{\rpv}_{\phi
K}$ is purely imaginary and introduces a new weak phase to the
decay amplitude of $B \to \phi K$ modes, which can cause non-zero
direct CP asymmetries, different from the SM prediction. In
contrast, $\bar {\cal A}^{\rpv}_{\eta' K}$ introduces no new phase
in this choice of $d^L_{222}$ and $d^R_{222}$. We find a
constructive contribution to the SM part of the amplitude for $B^+
\to \eta' K^+$ which helps to satisfy the experimental data.

\begin{table}
\caption{CP asymmetries in the decay modes $B^0 \to \phi K_S$ and
$B^0 \to \eta' K_S$. }
\begin{tabular}{|c|c|c|c|}
 $\sin(2 \tilde \beta)$ & Case 1 & Case 2 & experimental data\\ \hline
 $\sin(2 \tilde \beta)_{\phi K_S}$ & 0 & $-0.82$ &$-0.19^{+0.52}_{-0.50} \pm
 0.09$ (BaBar),  $-0.73 \pm 0.64 \pm 0.18$ (Belle) \\
 $\sin(2 \tilde \beta)_{\eta' K_S}$ & 0.73 & 0.72 &
 $0.76 \pm 0.36^{+0.05}_{-0.06}$ (Belle)
\end{tabular}
\end{table}

For $|\lambda^{\prime}_{322}| =|\lambda^{\prime}_{332}|
=|\lambda^{\prime}_{323}| =0.055$, $\tan\theta' = 0.52$, and $m_{\rm susy}
=200$ GeV, we find $\sin(2 \tilde \beta)_{\phi K_S} =0$ and $\sin(2 \tilde
\beta)_{\eta' K_S} =0.73$ for $\xi =0.45$, as shown in Table I. This result
agrees well with the recently reported experimental data. Figures 1 and 2
show the BR versus $\xi$ for $B^+ \to \phi K^+$ and $B^+ \to \eta' K^+$,
respectively.  In both figures, the shaded region represents the allowed
region by the experimental results and the solid line corresponds the SM
prediction.  The SM BR of $B^+ \to \eta' K^+$ is far below the experimental
lower bound.  Our result in this case is denoted by the dotted line, which
is clearly consistent with the experimental data: ${\cal B} (B^+ \to \eta'
K^+) = 69.3 \times 10^{-6}$ for $\xi =0.45$. In Table II, we estimate the
BRs and CP (rate) asymmetries ${\cal A}_{CP}$ for $B \to \eta^{(\prime)}
K^{(*)}$ and $B^{+} \to \phi K^{+}$, where ${\mathcal A}_{CP}$ is defined
by $ {\mathcal A}_{CP} = {{\mathcal B}(\bar b \rightarrow \bar f)
-{\mathcal B}(b \rightarrow f) \over {\mathcal B}(\bar b \rightarrow \bar
f) +{\mathcal B}(b \rightarrow f)}$ with $b$ and $f$ denoting $b$ quark and
a generic final state, respectively. The estimated BRs are well within the
experimental limits. The BR of $B^0\to\phi K^0$ has also been measured,
${\cal B}(B^0 \to\phi K^0) = (8.1^{+3.1}_{-2.5} \pm 0.8) \times
10^{-6}$\cite{babar}. The theoretical value for the BR of this mode is same
as that of $B^+ \to \phi K^+$ and is allowed by the experimental data.
Further, We note that ${\cal A}_{CP}$'s for $B^+  \to \phi K^+$ and $B^0
\to \eta K^{*0}$ are large and  have positive signs. This can be taken as a
prediction of this model.   It is hopeful that the CDF at Fermilab could
measure the CP asymmetry of $B^+ \to \phi K^+$ mode using their two track
trigger in the ongoing RUN II experiment\cite{teruki}.

So far we have assumed that the magnitudes and phases are same for
$d^L_{222}$ and $d^R_{222}$. We can assume them to be different
and still obtain good fits. For example, if we set $d^L_{222} =
k_L e^{-i \theta'_L}$ and $d^R_{222} = -k_R e^{i \theta'_R}$ and
choose: $k_L =1.0\times 10^{-8} $, $k_R = 8.0\times 10^{-9}$ and
$\tan \theta'_L=0.4$, $\tan \theta'_R =0.52$, we find $\sin(2
\tilde \beta)_{\phi K_S}=0.017$ and ${\cal B}(B^{\pm}\to\phi
K^{\pm})=7\times 10^{-6}$. The BRs of $B\to\eta{'}K$ modes remain
unchanged.

\begin{figure}
    \begin{center}
    \leavevmode
    \epsfysize=8.0cm
    \epsffile[75 160 575 630]{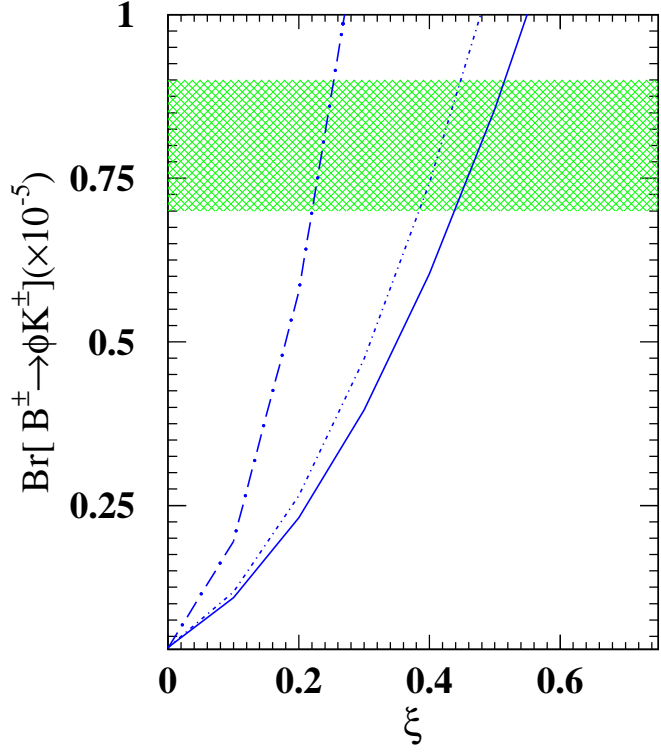}
    \vspace{2.0cm}
    \caption{\label{fig:fig1} BR vs $\xi$. The dotted and dot-dashed
    lines correspond to Case 1 and Case 2 respectively. The solid line
    corresponds to the SM (The SM BR is same for both cases).
    The shaded region is allowed by the experimental
    data.}
\end{center}
\end{figure}

\begin{table}
\caption{The branching ratios $({\cal B})$ and CP rate asymmetries
$({\cal A}_{CP})$ for $B \ra\eta^{(\prime)} K^{(*)}$ and $B \ra \phi K$.}
\begin{tabular}{|c|cc|cc|}
 & Case 1& & Case 2 &
\\ mode & ${\cal B} \times 10^{6}$ & ${\cal A}_{CP}$ & ${\cal B} \times 10^{6}$
& ${\cal A}_{CP}$
\\ \hline
$B^+ \to \eta^{\prime} K^+$ & 69.3 & $-0.01$ & 76.1 & $-0.01$
\\ $B^+  \to \eta K^{*+}$ &  27.9 & $-0.04$ & 35.2 & $-0.03$
\\ $B^0  \to \eta' K^0$ & 107.4 & 0.00 & 98.9 & 0.00
\\ $B^0  \to \eta K^{*0}$ & 20.5 & 0.71 & 11.7 & 0.15
\\ $B^+  \to \phi  K^+$ & 8.99 & $-0.21$ & 8.52 & $-0.25$
\end{tabular}
\end{table}

The other observed $B \to PP$ and $B \to VP$ decay modes without
$\eta^{(')}$ or $\phi$ in the final state, such as $B \to \pi \pi$, $B \to
\pi K$, $B \to \rho \pi$, $B \to \omega \pi$, and so on, are not affected
in this scenario, because $d^L_{222}$ and $d^R_{222}$ are relevant to the
process $b \to s \bar s s$ only. The estimated BRs for those modes by using
the above input values are consistent with the experimental data for
$\xi=0.45$ \cite{dko}.

\begin{figure}
    \begin{center}
    \leavevmode
    \epsfysize=8.0cm
    \epsffile[75 160 575 630]{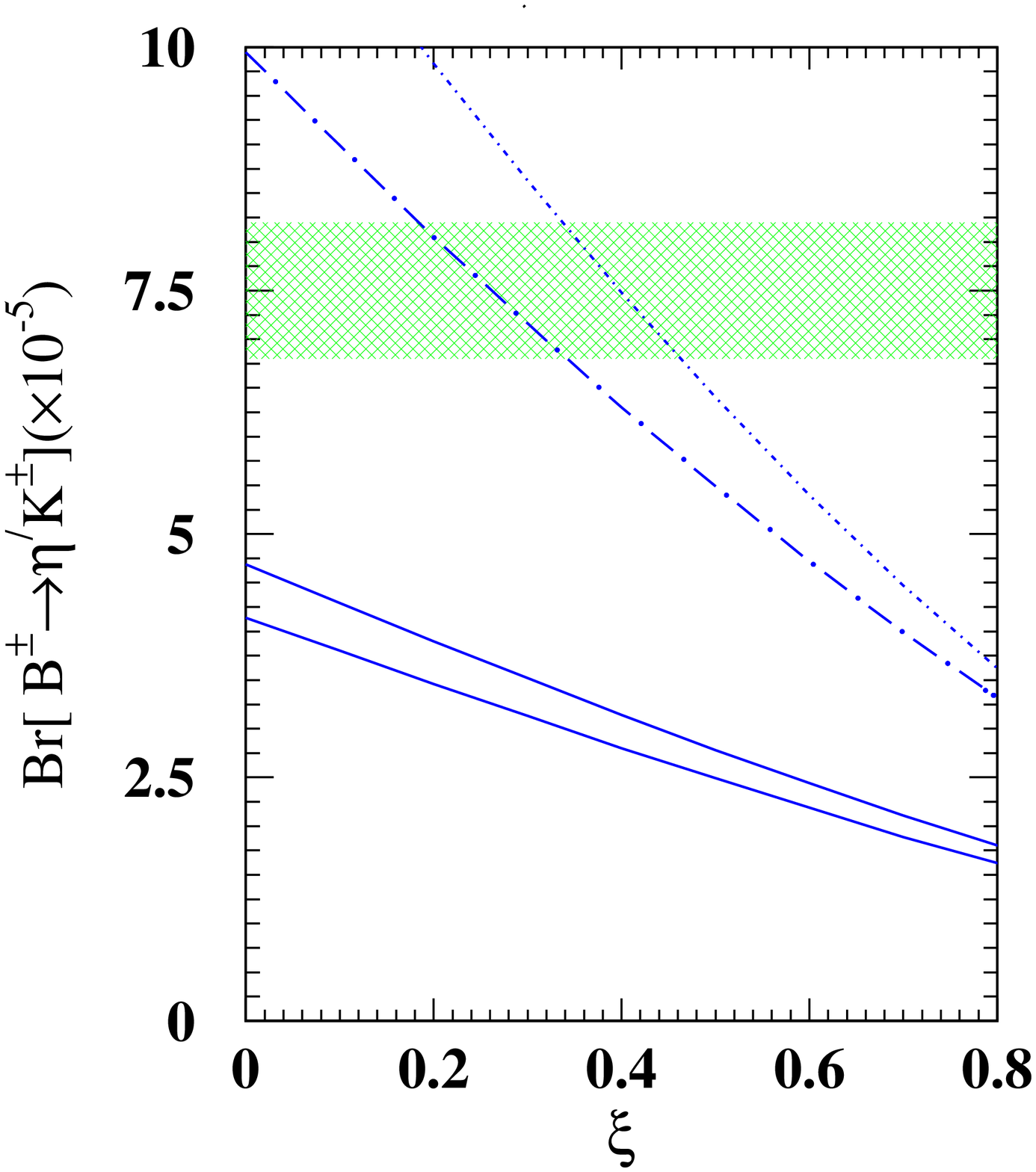}
    \vspace{2.0cm}
    \caption{\label{fig:fig2}  BR vs $\xi$. The dotted and dot-dashed
    lines correspond to Case 1 and Case 2 respectively. The solid lines
    correspond to the SM. The upper solid line is for Case 1 and the lower
    solid line is for Case 2. The shaded region is allowed by the experimental
    data.}
\end{center}
\end{figure}

{\bf Case 2:} In {Case 1}, we have generated the very small value for
$\sin(2 \tilde \beta)_{\phi K_S}$. Let us now try to generate a large
negative $\sin(2 \tilde \beta)_{\phi K_S}$, since  one experiment is
preferring a small negative value and the other experiment is preferring
large. We use the smaller value of $\gamma$ and $m_s$: $\gamma= 80^0$ and
$m_s~ ({\rm at}~ m_b ~{\rm scale}) =75$ MeV, keeping the other inputs
unchanged. We also use the same form of $d^L_{222}$ and $d^R_{222}$ as in
{Case 1}: the $\rpv$ part of the decay amplitude introduces a new weak
phase in the mode $B \to \phi K$, but no phase in the process $B^+ \to
\eta' K^+$.

In this case, we find that a consistent explanation of all the observed
data for $B \to PP$ and $B \to VP$ is possible for
$|\lambda^{\prime}_{322}| =|\lambda^{\prime}_{332}|
=|\lambda^{\prime}_{323}| =0.069$, $\tan\theta' = 2.8$, and $m_{\rm susy}
=200$ GeV. In this scenario, a large value of $\sin(2 \tilde \beta)$ with
the negative sign is possible for $B^0 \to \phi K_S$: $\sin(2 \tilde
\beta)_{\phi K_S} =-0.82$ for $\xi =0.25$, as shown in Table I.  The value
of $\sin(2 \tilde \beta)$ for $B^0 \to \eta' K_S$ is similar to that of
Case 1: $\sin(2 \tilde \beta)_{\eta' K_S} =0.72$ for $\xi =0.25$. The
estimated BRs for $B^+ \to \phi K^+$ and $B^+ \to \eta' K^+$ are denoted by
the dot-dashed lines in Figs. 1 and 2, respectively: ${\cal B} (B^+ \to
\phi K^+) = 8.52 \times 10^{-6}$ and ${\cal B} (B^+ \to \eta' K^+) = 76.1
\times 10^{-6}$ for $\xi =0.25$. The BRs of other $B\to PP,\,VP$ modes
satisfy the experimental data for $\xi =0.25$\cite{dko}.

In conclusion, we have shown that in $\rp$ violating  SUSY it is
possible to understand consistently both recently measured CP
asymmetry $\sin(2\beta)$ in $B^0 \to \phi K_S$ decay and the large
branching ratio of $B^{\pm}\rightarrow \eta' K^{\pm}$ decay, which
appear to be inconsistent with the SM prediction. Experimental
results imply that CP asymmetry for $B \to \phi K_s$ is different
from the SM value, but for $B \to \eta' K_s$ it is same as in the
SM. We have found that this tendency can be consistently
understood in the framework of this model.  We have searched for
possible parameter space and found that all the observed data for
$B \to PP$ and $B \to VP$ decays can be accommodated for certain
values of $\rpv$ couplings. Experimentally, these couplings can be
observed through the direct production of SUSY particles at
Tevatron (e.g., $\chi^{\pm}_1\chi^0_2$ production and their
subsequent decays to final states with multiple b quarks) and
other rare decays like $B_s \to \mu^{\pm} \mu^{\mp}$\cite{adkm}.

\vspace{1cm}
\centerline{\bf ACKNOWLEDGEMENTS}
\medskip

\noindent The work of C.S.K. was supported in part by  CHEP-SRC
Program, in part by Grant No. 20015-111-02-2 from DFG-KOSEF of the
KOSEF, in part by Grant No. R02-2002-000-00168-0 from BRP of the
KOSEF, in part by BK21 Program and Grant No. 2001-042-D00022 of
the KRF. The work of S.O. was supported in part by Grant No.
2001-042-D00022 of the KRF, and by the Japan Society for the
Promotion of Science (JSPS).


\end{document}